# THE COBE NORMALIZATION FOR STANDARD CDM

Emory F. Bunn, Douglas Scott & Martin White

Astronomy Department. University of California, Berkeley, CA 94720

Written 1994 October 15

## ABSTRACT

ABSTRACT: The *COBE* detection of microwave anisotropies provides the best way of fixing the amplitude of cosmological fluctuations on the largest scales. This normalization is usually given for an $n = 1$ spectrum, including only the anisotropy caused by the Sachs-Wolfe effect. This is certainly not a good approximation for a model containing any reasonable amount of baryonic matter. In fact, even tilted Sachs-Wolfe spectra are not a good fit to models like Cold Dark Matter (CDM). Here we normalize standard CDM (sCDM) to the 2-year *COBE* data, and quote the best amplitude in terms of the conventionally used measures of power. We also give normalizations for some specific variants of this standard model, and we indicate how the normalization depends on the assumed values of $n$, $\Omega_B$ and $H_0$. For sCDM we find $\langle Q \rangle = 19.9 \pm 1.5 \,\mu\mathrm{K}$, corresponding to $\sigma_8 = 1.34 \pm 0.10$, with the normalization at large scales being $B = (8.16 \pm 1.04) \times 10^5 (h^{-1}\,\mathrm{Mpc})^4$, and other numbers given in the Table. The measured rms temperature fluctuation smoothed on $10°$ is a little low relative to this normalization. This is mainly due to the low quadrupole in the data: when the quadrupole is removed, the measured value of $\sigma(10°)$ is quite consistent with the best-fitting $\langle Q \rangle$. The use of $\langle Q \rangle$ should be preferred over $\sigma(10°)$, when its value can be determined for a particular theory, since it makes full use of the data.

*Subject headings:* cosmic microwave background — large-scale structure of universe

## INTRODUCTION

During the 1980's, it was standard practice to normalize cosmological models at a scale of $\simeq 10 h^{-1}\,\mathrm{Mpc}$, using a quantity related to the clustering of galaxies (here the Hubble constant $H_0 = 100 h\,\mathrm{km\,s^{-1}\,Mpc^{-1}}$). Due to processing of the primordial spectrum, this method of normalization requires assumptions about both the equation of state for matter inside the horizon and the relationship between the observed structure and the underlying mass in the universe. After the *COBE* DMR detection of Cosmic Microwave Background (CMB) anisotropies (Smoot et al. 1992), it became possible to directly normalize the potential fluctuations at near-horizon scales, circumventing both of these problems with the 'conventional' normalization. To be specific, different authors have chosen to use one or other of the following *COBE*-derived quantities: (1) $\sigma(10°)$, the rms temperature fluctuation averaged over a $10°$ FWHM beam; (2) $Q_{\mathrm{rms-PS}}(n=1)$, the best-fitting amplitude for an $n = 1$ Harrison-Zel'dovich (HZ) spectrum, quoted at the quadrupole. We will henceforth refer to this quadrupole expectation value as '$\langle Q \rangle$'.

There are reasons for preferring one of these quantities over the other. First, $\sigma(10°)$ has the advantage of not depending on the assumption of a model in the fitting of the data, since it is purely an observationally-determined quantity. However, because of this, it does not provide the most accurate normalization for any specific theory. Furthermore, one must take great care to account properly for cosmic variance before using $\sigma(10°)$ to normalize a model (White et al. 1993). On the other hand, $\langle Q \rangle (n=1)$ gives the best amplitude for a *pure* HZ spectrum of CMB fluctuations, including only the Sachs-Wolfe (SW) effect (Sachs & Wolfe 1967). Although the SW effect is the main contributor to large-scale CMB anisotropy, other effects are not generally negligible, even on *COBE* scales. In particular, as shown in Figure 1, the sCDM model, with $\Omega_B = 0.05$ is not well approximated by a SW spectrum for any $n$. In other words, the low $\ell$ multipoles are not purely the SW potential fluctuations even on the angular scales probed by *COBE*.

Analysis of the first year of *COBE* data gave $\sigma(10°) = 30 \pm 5 \,\mu\mathrm{K}$ and $\langle Q \rangle (n=1) = 17 \pm 5 \,\mu\mathrm{K}$ (Smoot et al. 1992, Wright et al. 1994a); after the second year of data these became $\sigma(10°) = 30.5 \pm 2.7 \,\mu\mathrm{K}$ and $\langle Q \rangle (n=1) = 19.9 \pm 1.6 \,\mu\mathrm{K}$ (Bennett et al. 1994, Górski et al. 1994a, Wright et al. 1994b). With twice as much data there has been significant improvement in the error bars. But, confusingly, the best value for $\sigma(10°)$ has remained essentially unchanged, while $\langle Q \rangle$ has increased significantly. A simple calculation shows that $\sigma(10°)/\langle Q \rangle$ ought to be about 2 for $n = 1$, whereas the actual ratio is about 1.5. There is therefore a discrepancy of as much as 30% between these two choices of normalization, as also discussed by Banday et al. (1994).

This apparent inconsistency has led to some confusion about the exact normalization to use for a given theoretical model. This has implications for the question of how specific models fit the large-scale structure data (e.g. Efstathiou et al. 1992), what the required bias of galaxies is relative to dark matter, and whether particular models are consistent with smaller-scale CMB data (e.g. Górski et al. 1993, Dodelson & Jubas 1993, Srednicki et al. 1993, Bunn et al. 1994). It therefore seems important to normalize with the full accuracy available from the *COBE* data.

Other authors have tended to use the *COBE* data to fit $Q$ as a function of $n$ using the SW formula (e.g. Scaramella & Vittorio 1993, Seljak & Bertschinger 1993). One approximation for standard CDM is to take some value of $\langle Q \rangle$ for $n > 1$, say $n = 1.15$ (Bond 1993). However, even this is not an adequate representation of the sCDM model at small multipoles (Fig. 1). The best way to normalize a given model is to carry out a detailed fit to the data. The point of this letter is to perform this task for the totally 'vanilla-flavored' CDM, and for a range of minor deviations from it, so that it will be clear *exactly* what normalization to use in the future.

## COSMOLOGICAL MODELS

The cold dark matter model (see e.g. Ostriker 1993, Liddle & Lyth 1993) has become the 'straw man' model of structure formation. In this model $\Omega_0 = 1$, with a variable fraction $\Omega_B$ residing in baryons and the rest in massive (non-relativistic) dark matter. The initial fluctuations are assumed to be Gaussian distributed, adiabatic, scalar density fluctuations with an HZ spectrum on large scales, i.e. $P(k) \propto k^n$ with $n = 1$. The 'standard' CDM model has come to mean the one with $H_0 = 50\,\mathrm{km\,s^{-1}\,Mpc^{-1}}$, as a compromise between age and $H_0$ constraints, and with $\Omega_B = 0.05$, for consistency with the best central value from big-bang nucleosynthesis studies (Walker et al. 1991, Smith et al. 1993).

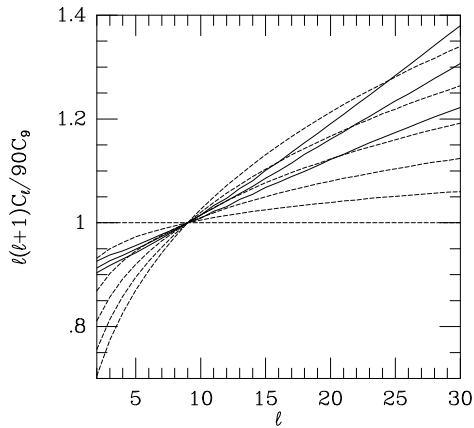

Fig. 1.— The solid lines are the CMB anisotropy multipoles for CDM models for $\Omega_B = 0.01$, 0.05 ('sCDM') and 0.10 (from bottom to top). The dashed lines are SW spectra for $n = 1.00$ to $n = 1.25$ in steps of 0.05, as computed using equation (4). The curves have all been normalized at $C_9$, which is approximately the 'pivot' point for the COBE data (Górski et al. 1994a). Note that the CDM spectra are not well-represented by any $n$.

Although it seems generally successful, one does not expect the sCDM model to be precisely the correct description of the universe, so it is worthwhile to consider a range of deviations from it. First one can allow for variations in the cosmological parameters $h$ and $\Omega_B$. On the scales probed by COBE the power spectrum is relatively insensitive to these changes (see below). Furthermore, the sCDM model assumes a 'flat' or HZ ($n = 1$) spectrum of density fluctuations. Since inflation generically predicts departures from the simple $n = 1$ form, we will quote normalizations for a range of $n$.

Should primordial gravitational waves (tensors) turn out to be important as well as the scalar fluctuations, their effect will be to lower the inferred $Q^2$ by roughly the fraction they contribute to $C_2$, i.e. $C_2^T/C_2^S$. This fraction is currently totally unknown, although it may be related to $n$ for specific models. For example, in 'extended' inflation it is roughly $7(1-n)$ (Crittenden et al. 1993), although in 'natural' inflation it is negligible even if $n < 1$ (Adams et al. 1992). Since including a tensor component also changes the shape of the $C_\ell$'s, we perform a fit to the data. The two specific possibilities mentioned above span the interesting range and we quote the normalization in both cases.

The addition of a component of Hot Dark Matter will cause a negligible effect at COBE scales, but will change the relevant normalization at cluster and galaxy scales. Models that have a cosmological constant and/or are open (including Baryonic Dark Matter models), as well as models containing topological defects, present additional difficulties, and we do not treat them in this letter.

### FITTING CDM TO THE COBE DATA

As usual in CMB studies we expand the temperature fluctuations $\Delta T/T(\theta,\phi)$ in spherical harmonics $\sum_{\ell m} a_{\ell m} Y_{\ell m}(\theta,\phi)$, and work in terms of the multipole moments $a_{\ell m}$. In general the predictions of a theory are expressed in terms of predictions for the $a_{\ell m}$. Using rotational symmetry $\langle a_{\ell m}^* a_{\ell' m'} \rangle \equiv C_\ell \delta_{\ell' \ell} \delta_{m' m}$, where the angled brackets represent an ensemble average. If the fluctuations are Gaussian, the predictions are fully specified by giving these $C_\ell$'s. We have calculated the $C_\ell$'s using a Boltzmann code, and have compared them to other calculations (N. Sugiyama, private communication; S. Dodelson, private communication), finding that agreement between the models is better than $\simeq 3\%$ (1.5% in temperature) over the range $\ell = 2$ to 2000.

We fit to the data using the 'eigenmodes of the signal-to-noise' approach discussed in Bunn & Sugiyama (1994), and similar to the approach of Bond (1994a, 1994b). We expand the map in terms of a set of basis functions, and use the coefficients of this expansion to compute likelihoods for each model we wish to test. The basis functions are chosen to have the maximum possible rejection power for incorrect models. (Note that we do not compute new eigenmodes for each new model we test: the same functions are used throughout.) We keep only the 400 most significant modes in the expansion; retaining more modes does not significantly increase the rejection power (Bunn & Sugiyama 1994). We use the weighted average of the 31, 53 & 90 GHz two-year COBE maps, with a galactic cut at $|b| < 20°$, and with the monopole and dipole removed. We have checked that other reasonable map combinations (e.g., using only the 53 and 90 GHz maps) give consistent answers. Our analysis is insensitive to the monopole and dipole in the maps, since we marginalize over these quantities; nonetheless, we remove a best-fit monopole and dipole from the maps. We have also performed Monte Carlo simulations to test whether this technique gives an unbiased estimate of $\langle Q \rangle$. We generated simulated data sets for an HZ power spectrum with $\langle Q \rangle = 21\,\mu\mathrm{K}$, and for an sCDM power spectrum with $\langle Q \rangle = 20\,\mu\mathrm{K}$. We recovered a mean $\langle Q \rangle$ which differed by less than $0.1\,\mu\mathrm{K}$.

We have applied this technique to determine the likelihood function for $\langle Q \rangle$ for a number of models. For sCDM we find $\langle Q \rangle = 19.9 \pm 1.5\,\mu\mathrm{K}$ with the quadrupole included, and $\langle Q \rangle = 20.3 \pm 1.5\,\mu\mathrm{K}$ with the quadrupole removed. In terms of power $\langle Q^2 \rangle = 403 \pm 61\,(\mu\mathrm{K})^2$, with the quadrupole and $\langle Q^2 \rangle = 418 \pm 64\,(\mu\mathrm{K})^2$ without. For a tilted CDM model (with no tensors), the normalization is

$$\langle Q \rangle = 21 \exp[0.69(1-n)]\,\mu\mathrm{K} \qquad (1)$$

to an accuracy better than 0.5% for $0.5 < n < 2$, while best fit is $(\langle Q \rangle, n) = (17.1\,\mu\mathrm{K}, 1.3)$ [and $(\langle Q \rangle, n) = (18.6\,\mu\mathrm{K}, 1.2)$ without the quadrupole]. The value of $Q$ also depends weakly on $\Omega_B$ and is virtually independent of $h$, as indicated in the Table. Both $\Omega_B$ and $h$ additionally affect the transfer function and hence the smaller scale normalizations. Our results are about $1\,\mu\mathrm{K}$ higher than those in Górski et al. (1994a). This appears to be due to the fact that we use the publically available sky maps rather than those used by the COBE group: when the Górski et al. analysis is performed on the publically available maps, the results agree with ours to with 0.5 $\mu\mathrm{K}$. (Górski, private communication; Górski et al. 1994b). In any case, the discrepancy is well within the uncertainties.

Including tensor modes reduces the goodness of fit to the data, since it increases the predicted quadrupole. With $C_2^T/C_2^S = 7(1-n)$ we find that the best-fitting values of $\langle Q \rangle = (\langle Q \rangle_T^2 + \langle Q \rangle_S^2)^{1/2}$ are 22.9, 25.2 and $27.5\,\mu\mathrm{K}$ for CDM models with $n = 0.9$, 0.8 and 0.7, respectively.

### THE 10° VARIANCE

Given any set of $C_\ell$'s one can predict the rms fluctuation, smoothed over 10°, through

$$\langle \sigma^2(10°) \rangle = \frac{1}{4\pi} \sum_{\ell=2}^{\infty} (2\ell+1) C_\ell W_\ell, \qquad (2)$$

where $W_\ell$ is the window function (Wright et al. 1994a). Thus it is possible to use $\sigma(10°)$ as a means of quoting the normalization. Using a simple $10°$ FWHM Gaussian for the window function leads to $\sigma(10°) = 1.99 \langle Q \rangle$ for a pure $n = 1$ model. Using the window function and correction for the smoothing of Wright et al. (1994a) one has $\sigma/\langle Q \rangle = 1.91$; the corresponding ratio for sCDM is 1.95. The data give us $\sigma/\langle Q \rangle = 1.63$, which we find in less than 7% of a set of simulated *COBE* skies made with the sCDM power spectrum. Using $n = 1.5$ CDM instead of sCDM alleviates the problem: the predicted $\sigma/\langle Q \rangle$ is 2.32, while the data give $\sigma/\langle Q \rangle = 2.31$.

However, if we remove the quadrupole from the data, then $\sigma(10°)$ and $\langle Q \rangle$ are quite consistent with each other. The ratio found in the data after removing the quadrupole is $\sigma/\langle Q \rangle = 1.54$, which agrees well with the prediction of 1.69 for sCDM. The lesson here is that it is better to normalize a model directly to the data, if it is available, rather than using the quoted $10°$ variance. If you do use $\sigma(10°)$ then you certainly need to include the effects of the beam of Wright et al. (1994a), which lowers $\sigma/\langle Q \rangle$ by $\simeq 5\%$. The remaining 'discrepancy' is due to the quadrupole being significantly lower than expected for flat models (see Bennett et al. 1994).

OTHER WAYS TO QUOTE THE NORMALIZATION.

Let us take the spectrum of primordial fluctuations to be a power law in comoving wavenumber $k$. If we assume that temperature fluctuations at large scales arise from the SW effect, $\Delta T/T = 1/3 \Phi$, the spectrum of radiation fluctuations looks like

$$\mathcal{P}(k) \equiv A (k\eta_0)^{n-1}. \qquad (3)$$

Here $\eta_0 \simeq 3t_0 = 2H_0^{-1}$ (for $\Omega_0 = 1$) is the conformal time today, with scale factor normalized to unity and $A$ is one way of quoting the amplitude for scalar perturbations. It is related to $\varepsilon_H$, the dimensionless amplitude of matter fluctuations at horizon crossing, through $\varepsilon_H^2 = (4/\pi)A$.

Assuming the SW result above, we can write the average over universes of the moments of the temperature anisotropy as

$$C_\ell = 2^n \pi^2 A \frac{\Gamma(3-n)\Gamma\left(\ell + \frac{n-1}{2}\right)}{\Gamma^2\left(\frac{4-n}{2}\right)\Gamma\left(\ell + \frac{5-n}{2}\right)} \qquad (4)$$

(see e.g. Peebles 1982, Abbott & Wise 1984, Bond & Efstathiou 1987, White et al. 1994). For the special case of $n = 1$, we have $C_2/A = 4\pi/3$ and $C_\ell^{-1} \propto \ell(\ell+1)$. Often one quotes not $C_2$ but

$$\langle Q \rangle = T_0 \left(\frac{5 C_2}{4\pi}\right)^{1/2} = T_0 \left(\frac{5}{3}A\right)^{1/2}, \qquad (5)$$

which is the convention used in, e.g., Smoot et al. (1992), and where we use $T_0 = 2.726$K (Mather et al. 1994). However, as has been emphasized before (Bond 1993), the $C_\ell$'s for sCDM can depart significantly from equation (4) even for $\ell \sim 10$. In fact they are not well fit by the SW formula for any value of $n$ (Fig. 1). Note that for $n > 1$ the low-$\ell$ multipoles are quite suppressed relative to $\ell \sim 10$, which goes some way toward explaining why larger $n$ are preferred by the *COBE* data.

Another common normalization convention is to define the *matter* power spectrum as

$$P(k) \equiv B k^n T^2(k), \qquad (6)$$

where the transfer function $T(k) \simeq 1$ on large scales. This means that the dimensions of $B$ will depend on $n$ (and will be Length[4] for $n = 1$). If the fluctuations arise purely from the SW effect, then $B$ and $A$ are simply related, so for $n = 1$ (White et al. 1994)

$$P(k) = 2\pi^2 \eta_0^4 A k T_m^2(k) \qquad (7)$$
$$\simeq 2.5 \times 10^{16} A (k/h \, \text{Mpc}^{-1}) T^2(k) \, (h^{-1} \text{Mpc})^3. \qquad (8)$$

So for an $n = 1$ primordial spectrum $B = 2\pi^2 \eta_0^4 A = (6\pi^2/5) \eta_0^4 \langle Q^2 \rangle / T_0^2$. The assumption of pure SW fluctuations is not exact (see Fig. 1), although when we normalize to the quadrupole it is approximately correct. A Boltzmann calculation in which the photon and dark matter perturbations are explicitly evolved in time shows this relation holds to an accuracy of $\lesssim 4\%$. There is better agreement for higher values of $h$, which moves last scattering further into the matter dominated regime. For example, if $h = 1$ the SW formula (for $C_2$) is good to $\lesssim 1\%$. We have used the ratio of matter to radiation normalizations from the Boltzmann calculation where appropriate.

For CDM one conventionally uses a parameterized transfer function, $T(k) = \left[1 + \left(ak + (bk)^{3/2} + (ck)^2\right)^\nu\right]^{-1/\nu}$ (Efstathiou 1990) with $a = (6.4/\Gamma)h^{-1}$Mpc, $b = (3/\Gamma)h^{-1}$Mpc, $c = (1.7/\Gamma)h^{-1}$Mpc, $\nu = 1.13$ and the shape parameter $\Gamma \simeq \Omega_0 h$. We use a calculation of the transfer function for our specific model ($h = 0.5$, $\Omega_B = 0.05$) which fits the above form with $\Gamma = 0.48$. We find that variations in the transfer function with $\Omega_B$ can lead to 10% changes in small scale power.

Large-scale flows also provide a measure of the power spectrum: the variance of the velocity field in spheres of radius $r$, $V_{\rm rms}^2(r)$ can be expressed as an integral over the power spectrum (e.g., Peebles 1993). This tends to probe scales similar to the degree-scale CMB experiments. Bertschinger et al. (1990) estimated the 3D velocity dispersion of galaxies within spheres of radius $40h^{-1}$Mpc and $60h^{-1}$Mpc, after smoothing with a Gaussian filter on $12h^{-1}$Mpc scales. So we also quote the normalization in terms of the quantities $V_{40}$ and $V_{60}$, corresponding to the above procedure.

On smaller scales, associated with clusters of galaxies, one conventionally quotes $J_3(10h^{-1} \text{Mpc})$ (e.g. Peebles 1993), and the 'bias' $b_\rho$ or variance of the density field in spheres of $8h^{-1}$Mpc radius $\sigma_8$, defined through

$$b_\rho^{-2} \equiv \sigma_8^2 \equiv \int_0^\infty \frac{dk}{k} A(k\eta_0)^{n+3} T^2(k) \left(\frac{3 j_1(kr)}{kr}\right)^2, \qquad (9)$$

where $r = 8 h^{-1}$Mpc. The variance of galaxies, possibly biased relative to the matter ($\delta_{\rm gal} = b \delta_\rho$), is roughly unity on a scale of $8h^{-1}$Mpc (Davis & Peebles 1983, Loveday et al. 1992). For sCDM the *COBE* best-fit gives $\sigma_8 \simeq 1.3$, i.e. a relatively unbiased model. However this depends on the values of $\Omega_0$, $h$ etc. that are adopted, as indicated in the Table.

CONCLUSIONS

The quality of the data used to test theories of large-scale structure has improved markedly in recent years. There is even some hope that the CMB will soon allow us to make quantitative estimates of important cosmological parameters (Bond et al. 1994, Scott et al. 1994). In particular, the *COBE* data are of sufficiently high quality that some care is required to determine the proper *COBE* normalization for any particular theory. When normalizing a theory such as sCDM, it is not correct to assume a pure SW HZ power spectrum, or even an $n = 1.15$ SW power spectrum. For sCDM and several of its popular variants, the normalizations in the Table are to be preferred.

There is an apparent discrepancy between the normalizations in the Table and the measured temperature anisotropy on an angular scale of 10°: the value of the rms determined from the sky maps is approximately 20% lower than the ensemble-average values $\langle \sigma(10°) \rangle$ in the Table. The agreement is improved by using the non-Gaussian COBE beam. Even so, Monte Carlo simulations show that this discrepancy is significant at a confidence level of $\simeq$ 93%, but that removing the quadrupole from both the data and the simulations greatly alleviates the problem. Since the value of $\sigma(10°)$ is so sensitive to the quadrupole, and since there is some reason to suspect that the quadrupole may be contaminated, $\sigma(10°)$ is not a good quantity to use in normalizing models. In any case, an estimate of $\langle Q \rangle$ derived from a detailed fit to a model is to be preferred over $\sigma(10°)$, since the fit to $\langle Q \rangle$ makes use of all of the information in the data.

We would like to thank K. Górski, W. Hu and J. Silk for helpful discussions. The COBE data sets were developed by the NASA Goddard Space Flight Center under the guidance of the COBE Science Working Group and were provided by the NSSDC. This work was supported by grants from NASA, the NSF and the TNRLC.

REFERENCES

Abbott, L. F., & Wise, M. B., 1984, ApJ, **282**, L47
Adams, F. C., Bond, J. R., Freese, K., Freeman, J. A. & Olinto, A., 1992, Phys. Rev. D, **47**, 426
Banday, A. J., et al. 1994, ApJ, submitted
Bennett, C. L., et al. 1994, ApJ, in press
Bertschinger, E., et al. 1990, ApJ, **364**, 370
Bond, J. R., 1993, In Proceedings of the IUCAA Dedication Ceremonies, ed. T Padmanabhan, New York: John Wiley & Sons, in press
Bond, J. R., 1994a, Astrophys. Lett. & Commun., in press
Bond, J. R., 1994b, preprint, CITA
Bond J. R., Crittenden R., Davis R. L., Efstathiou G., Steinhardt P. J. 1994, Phys. Rev. Lett., **72**, 13
Bond, J. R. & Efstathiou, G., 1987, MNRAS, **226**, 655
Bunn, E. F. & Sugiyama, N., 1994, ApJ, submitted
Bunn, E. F., White, M., Srednicki, M. & Scott, D., 1994, ApJ, **429**, 1
Crittenden, R., Bond, J. R., Davis R. L., Efstathiou, G., & Steinhardt, P. J., 1993, Phys. Rev. Lett., **71**, 324
Davis, M. & Peebles, P. J. E., 1983, ApJ, **267**, 465
Dodelson, S. & Jubas, J. 1993, Phys.Rev.Lett., **70**, 2224
Efstathiou G. 1990, In Physics of the Early Universe: Proceedings of the 36th Scottish Universities Summer School in Physics, ed. J. A. Peacock, A. E. Heavens, A. T. Davies, p.361. New York: Adam Hilger
Efstathiou, G., Bond, J. R. & White, S. D. M., 1992, MNRAS, **258**, 1p
Górski, K. M., Stompor, R., & Juszkiewicz, R., 1993, ApJ, **410**, L1
Górski, K. M., et al. 1994a, ApJ, **430**, L89
Górski, K. M., et al., 1994b, in preparation,
Holtzman, J. A., 1989, ApJS, **71**, 1
Hu, W. & Sugiyama, N., 1994, Preprint, CfPA-TH-94-34
Liddle, A. R. & Lyth, D. H., 1993, Phys. Rep., **231**, 1
Loveday, J., Efstathiou, G., Peterson, B. A., & Maddox, S. J., 1992, ApJ, **400**, L43
Mather, J. C., et al. 1994, ApJ, **420**, 439
Ostriker, J. P., 1993, ARAA, **31**, 689
Peebles, P. J. E., 1982, ApJ, **263**, L1
Peebles, P. J. E., 1993, Principles of Physical Cosmology, Princeton NJ: Princeton University Press
Sachs, R. K. & Wolfe, A. M., 1967, ApJ, **147**, 73
Scaramella, R. & Vittorio, N., 1993, MNRAS, **263**, L17
Scott, D., Silk, J. & White, M., 1994, Preprint, Berkeley
Seljak, U. & Bertschinger, E., 1993, ApJ, **417**, L9
Smith, M. S., Kawano, L. H. & Malaney, R. A., 1993, ApJS, **85**, 219
Smoot, G. F., et al. 1992, ApJ, **396**, L1
Srednicki M., White M., Scott D., Bunn E. F., 1993, Phys. Rev. Lett., **71**, 3747
Walker, P. N., Steigman, G., Schramm, D. N., Olive, K. A. & Kang, H.-S., 1991, ApJ, **376**, 51
White, M., Krauss, L. & Silk, J., 1993, ApJ, **418**, 535
White, M., Scott, D. & Silk, J., 1994, ARAA, **32**, 319
Wright, E. L., et al. 1994a, ApJ, **420**, 1
Wright, E. L., Smoot, G. F., Bennett, C. L. & Lubin, P. M., 1994b, ApJ, in press

| Model | $10^{11}A$ | $10^{-5}B$ $(h^{-1}\text{Mpc})^{3+n}$ | $10^{10}C_2$ | $\langle Q \rangle$ ($\mu$K) | $\langle \sigma(10°) \rangle$ ($\mu$K) | $V_{60}$ (km s$^{-1}$) | $V_{40}$ (km s$^{-1}$) | $J_3(10)$ ($h^{-1}$Mpc)$^3$ | $\sigma_8$ |
|---|---|---|---|---|---|---|---|---|---|
| S-W $n=1$ | 3.65 | 8.84 | 1.53 | 21.1 | 42.0 | — | — | — | — |
| S-W $n=1.15$ | 2.48 | 22.17 | 1.24 | 19.0 | 40.1 | — | — | — | — |
| sCDM | 3.26 | 8.16 | 1.36 | 19.9 | 40.6 | 354 | 442 | 543 | 1.34 |
| $\Omega_B = 1\%$ | 3.32 | 8.34 | 1.39 | 20.1 | 40.8 | 360 | 453 | 639 | 1.45 |
| $\Omega_B = 10\%$ | 3.21 | 8.03 | 1.34 | 19.8 | 40.4 | 349 | 433 | 450 | 1.22 |
| $h = 0.3$ | 3.23 | 8.05 | 1.35 | 19.8 | 40.6 | 298 | 357 | 149 | 0.67 |
| tCDM (S) | 4.19 | 4.26 | 1.57 | 21.4 | 41.0 | 305 | 376 | 336 | 1.03 |
| tCDM (S+T) | 2.85 | 2.90 | 1.81 | 22.9 | 43.6 | 252 | 310 | 229 | 0.84 |
| MDM | 3.26 | 8.16 | 1.36 | 19.9 | 40.6 | 370 | 463 | 329 | 0.97 |

Table.— COBE normalizations for standard Cold Dark Matter (sCDM) and several of its variants. The first three normalizations use our best fit for $\langle Q^2 \rangle$ (with typical '1$\sigma$' uncertainties $\simeq$ 15%), while the others are proportional to $\langle Q \rangle$ (error $\simeq$ 7.5%). The first two models were normalized using equation (4) for the angular power spectrum, neglecting everything but the Sachs-Wolfe effect. The third model is sCDM. The next three models show the effect of varying the baryon fraction (at fixed $h = 0.5$) and the Hubble constant (with $\Omega_B h^2 = 0.0125$). tCDM is a tilted CDM model, with $n = 0.9$, given for the case with scalars only (S) and with a component of tensors added (S+T). MDM is a mixed dark matter model containing 30% hot and 69% cold dark matter, taken from Holtzman (1989). The power normalizations $A$ and $B$ are defined in equations (3) and (6) respectively. $C_2$ is the value of the angular power spectrum at $\ell = 2$. We have emphasized that the derived values of $Q$ and $\sigma$ are expectation values – this is strictly true of the other quantities also. Here by $\sigma(10°)$ we mean equation (2) smoothed with a 10° FWHM Gaussian. $V_{60}$ and $V_{40}$ are the predicted rms velocities in spheres of radius $60h^{-1}$Mpc and $40h^{-1}$Mpc after smoothing with a Gaussian of width $12h^{-1}$Mpc. The $J_3(10)$ and $\sigma_8$ normalizations are defined in the text.